\documentclass[nofootinbib,superscriptaddress, a4paper, twocolumn, floatfix]{revtex4-2}
\usepackage[english]{babel}
\usepackage[utf8]{inputenc}
\usepackage[colorinlistoftodos, color=green!40, prependcaption]{todonotes}
\usepackage{amsthm}
\usepackage{mathtools}
\usepackage{physics}
\usepackage{braket}
\usepackage{listings}
\usepackage{xcolor}
\usepackage{graphicx}
\usepackage[left=23mm,right=13mm,top=35mm,columnsep=15pt]{geometry} 
\usepackage{adjustbox}
\usepackage{placeins}
\usepackage[T1]{fontenc}
\usepackage{lipsum}
\usepackage{csquotes}
\usepackage[pdftex, pdftitle={Article}, pdfauthor={Author}]{hyperref} 
\usepackage{booktabs}
\usepackage{amssymb}
\usepackage{bm}
\usepackage[ruled,vlined]{algorithm2e}
\usepackage[caption=false]{subfig}
\usepackage{stmaryrd}

\usepackage[ruled,vlined]{algorithm2e}

\graphicspath{{Figures/}} 

\usepackage{mathtools}
\usepackage{nameref}
\usepackage{varioref}
\usepackage{hyperref}
\usepackage{cleveref}
\crefname{equation}{Eq.}{Eqs.}
\Crefname{equation}{Equation}{Equations}
\crefname{figure}{Fig.}{Figs.}
\Crefname{figure}{Figure}{Figures}
\crefname{section}{Sec.}{Sects.}
\Crefname{section}{Section}{Sections}
\crefname{table}{Table}{Tables}
\crefname{appsec}{Appendix}{Appendices}
\Crefname{paragraph}{Section}{Sections}
\crefname{algorithm}{Algorithm}{Algorithms}

\newcommand{\hH}{\hat{H}}

\newcommand{\hI}{\hat{I}}

\newcommand{\hU}{\hat{U}}

\newcommand{\mcL}{\mathcal{L}}

\newcommand{\Hp}{\hH_{\text{prob}}}

\newcommand{\hX}{\hat{X}}
\newcommand{\hY}{\hat{Y}}
\newcommand{\hZ}{\hat{Z}}

\newcommand{\ie}{\textit{i.e.} }
\newcommand{\eg}{\textit{e.g.} }

\newcommand*{\mathcolor}{}
\def\mathcolor#1#{\mathcoloraux{#1}}
\newcommand*{\mathcoloraux}[3]{%
  \protect\leavevmode
  \begingroup
    \color#1{#2}#3%
  \endgroup
}

\begin{document}
\title{Exploring the role of parameters in variational quantum algorithms}

\author{Abhinav Anand}
\thanks{These authors contributed equally to this work}
\email[]{abhinav.anand@mail.utoronto.ca}
\affiliation{Chemical Physics Theory Group, Department of Chemistry, University of Toronto, Canada.}

\author{Sumner Alperin-Lea}
\thanks{These authors contributed equally to this work}
\email[]{sumner.alperin@mail.utoronto.ca}
\affiliation{Chemical Physics Theory Group, Department of Chemistry, University of Toronto, Canada.}

\author{Alexandre Choquette}
\thanks{These authors contributed equally to this work}
\email[]{alexandrecpoitevin@gmail.com}
\affiliation{Department of Computer Science, University of Toronto, Canada.}
\affiliation{Vector Institute for Artificial Intelligence, Toronto, Canada.}

\author{Al\'{a}n Aspuru-Guzik}
\email[]{aspuru@utoronto.ca}
\affiliation{Chemical Physics Theory Group, Department of Chemistry, University of Toronto, Canada.}
\affiliation{Department of Computer Science, University of Toronto, Canada.}
\affiliation{Department of Chemical Engineering and Applied Chemistry,  University of Toronto, Canada.}
\affiliation{Department of Materials Science and Engineering, University of Toronto, Canada.}
\affiliation{Vector Institute for Artificial Intelligence, Toronto, Canada.}
\affiliation{Canadian  Institute  for  Advanced  Research  (CIFAR)  Lebovic  Fellow,  Toronto,  Canada.}
\date{\today}

\begin{abstract}
        In this work, we introduce a quantum-control-inspired method for the characterization of variational quantum circuits using the rank of the dynamical Lie algebra associated with the hermitian generator(s) of the individual layers. Layer-based architectures in variational algorithms for the calculation of ground-state energies of physical systems are taken as the focus of this exploration. A promising connection is found between the Lie rank, the accuracy of calculated energies, and the requisite depth to attain target states via a given circuit architecture, even when using a lot of parameters which is appreciably below the number of separate terms in the generators. As the cost of calculating the dynamical Lie rank via an iterative process grows exponentially with the number of qubits in the circuit and therefore becomes prohibitive quickly, reliable approximations thereto are desirable.
        The rapidity of the increase of the dynamical Lie rank in the first few iterations of the calculation is found to be a viable (lower bound) proxy for the full calculation, balancing accuracy and computational expense. We, therefore, propose the dynamical Lie rank and proxies thereof as a useful design metric for layer-structured quantum circuits in variational algorithms.
\end{abstract}

\maketitle

\section{Introduction}
Variational quantum algorithms (VQA)~\cite{mcclean2016theory, bharti2022noisy, cerezo2021variational, anand2022quantum} are believed to be the most likely candidate to furnish practical applications for near-term quantum computers.
These hybrid quantum-classical routines combine the iterative execution of parametrized quantum circuits (PQC) with a classical optimization routine employed to minimize some cost function based on quantum observables. 
Without loss of generality, this cost function can be considered an approximation to the ground state energy of the problem Hamiltonian.

Both circuit design and optimization routine selection have been performed in a broadly heuristic manner. 
In recent years, exploration of the expressibility of PQCs  has furnished quantum algorithmists with new ways of characterizing variational circuits on theoretical grounds, including techniques based on the Meyer-Wallach Q measure~\cite{sim2019expressibility}, the Fisher information matrix~\cite{abbas2021power}, and memory capacity~\cite{wright2020capacity}. Wide classes of circuits have been investigated~\cite{nakaji2021expressibility,haug2021capacity} .

Futheremore, connections between quantum controllability and VQAs have been studied in varied contexts~\cite{magann2020pulses,lloyd2018quantum,morales2020universality,mbeng_quantum_2019,akshay_reachability_2020,choquette2020quantum}.
For instance, \citet{magann2020pulses}~pointed out that both quantum control and VQAs can be seen as quantum-classical optimization tasks.
This facilitates the use of the quantum control language in the VQA setting.
Moreover, \citet{lloyd2018quantum} used Lie algebraic arguments to show that QAOA is computationally universal, a proof that was formalized later by \citet{morales2020universality}.

A notion of reachability was proposed in Ref.~\cite{akshay_reachability_2020} to analyze the capacity of a given parametrized quantum circuit to represent a quantum state.
They applied the QAOA approach from Ref.~\cite{morales2020universality} to study MAX-2-SAT and MAX-3-SAT optimization problem~\cite{akshay2021reachability}.
The authors found that beyond a critical constraints-to-variables ratio in the problem setting, the algorithm exhibits reachability deficits, that is the QAOA ansatz does not cover a sufficiently large portion of the Hilbert space to contain the target state. 
Similar reachability issues have also been found in the variational Grover search problem~\cite{akshay_reachability_2020}.
This indicates that a better understanding of the Hilbert space coverage of variational ans\"atze would be beneficial. For instance, one could avoid reachability deficits by introducing symmetry-breaking terms or adopting a different parametrization of the PQCs that increases the dimension of the solution space.

In this work, we seek to introduce a notion of reachability for a class of PQCs by using its connection with quantum optimal control theory.
First, we present the similarities between the definition of a Hamiltonian-based PQC and the exponential map of the Lie algebra spanned by said Hamiltonian.
Next, we present a way to use the controllability of a system to determine the set of reachable states of the PQC.
Finally, we carry out numerical experiments to verify our method using the variational quantum eigensolver (VQE) framework~\cite{peruzzo2014variational}.
The results from our numerical experiments suggest that the dynamical Lie algebra can be used as a tool to correlate the parameterization of PQCs to reachability issues\cite{akshay2020reachability}.

The rest of the paper is organized as follows: we provide some background information about controllability and the Lie algebra formalism in \cref{section:prelims}, the details of our method in \cref{sec:method}, the result from numerical simulations in \cref{sec:simulations}, and concluding remarks in \cref{sec:conclusion}. 

\section{Preliminaries}\label{section:prelims}

\subsection{Controllability of quantum systems}

In this section, we seek to formalize the notion of controllability of controlled quantum systems described by the time-dependent Hamiltonian
\begin{equation}\label{eq:SE}
    \hH(t)=\sum_k c_k(t)\hH_k. 
\end{equation}
The Schr\"odinger equation of the control problem can be written as the system of differential equations:
\begin{equation}\label{eq: U SE}
        \dot\hU(t) = -i\left[ \sum_k c_k(t)\hH_k \right] \hU(t), \ \ \ \hU(0)=I
\end{equation}
where the solution to \cref{eq: U SE} is the unitary~$\hU(t)$, which propagates pure states through time as~$\ket{\psi(t)}=\hU(t)\ket{\psi(0)}$
for different set of controls $\{ c_k(t) \}$.
The system~$\hH(t)$ is said to be fully controllable if the set of possible operators that can be obtained from solving~\cref{eq: U SE} for all~$\{ c_k(t) \}$ is the set of all the unitary matrices. 
In this case, for any initial state~$\ket{\psi(0)}$, there exists a set of controls~$\{c_k(t)\}$ and a time~$T>0$ for which the state~$\ket{\psi(T)}$ can be mapped to any target state of the full Hilbert space.
We begin by detailing what a reachable set consists of.

The reachable set at a time~$T>0$ for system described by \cref{eq:SE},~$\mathcal{R}(T)$, is the set of all unitary matrices~$\bar U$ that one can obtain using the controls~$\{c_k\} \in \bar{\mathcal{C}}$.
In other words,~$\bar U = \hU (T,\{c_k\} )$. 
Here, we take~$\bar{\mathcal{C}}$ to be the set of piecewise constant functions as the time is often discretized when solving the Schr\"odinger equation.
Then, the reachable set at all times is defined as
\begin{equation}
        \mathcal{R} = \cup_{T\geq 0} \mathcal{R}(T)
\end{equation}
and comprises all possible states in the Hilbert space that system described by \cref{eq:SE} can reach.
$\mathcal{R}$ is therefore directly connected to the controllability of the system.

\subsection{Dynamical Lie algebra and controllability}

There exists an elegant connection between the set of reachable states of a system,~$\mathcal{R}$, and its dynamical Lie algebra,~$\mathcal{L}$.
This connection is well summarized by Theorem 3.2.1 of~\citet{d2007introduction} which states that

\emph{Theorem 1} \textit{The set of reachable states for a system $\sum_k c_k(t) \hH_k$ is the connected Lie group associated with the dynamical Lie algebra $\mathcal{L}$ generated by} $\text{span}_{\{c_k\} \in \bar{\mathcal{C}}}\{i\hH_k\}$. \textit{In short,}
\begin{equation}
        \mathcal{R} = e^{\mathcal{L}},
\end{equation}
\textit{where the Lie group $e^{\mathcal{L}}$ is the exponential map of the Lie algebra~$\mathcal{L}$, defined as }
\begin{equation}
    e^{\mathcal{L}}\equiv \{ e^{A_1}e^{A_2}\cdots e^{A_m}: A_1,A_2,\dots,A_m \in \mathcal{L} \}.
\end{equation}

Theorem 1 means that once the dynamical Lie algebra,~$\mcL$, of a system is known, one straightforwardly has access to its set of reachable states, and one has therefore a clear notion of the controllability of the system.
In fact,~$\mcL$ gives a basis of the vector space in which the controlled system is defined,~\ie, the size of the Hilbert subspace reachable by tuning the controls.
The quantity~$\text{dim}(\mathcal{L})$ is often referred to as the~\textbf{Lie rank criterion} and gives the dimension of that vector space. 

The system is fully controllable if $\text{dim}(\mathcal{L}) = n^2 = \text{dim}(u(n))$, where $n$ is the size of the full Hilbert space and $u(n)$ is the Lie algebra of skew-Hermitian $n\times n$ matrices. 
This is equivalent to $\mcL = u(n)$ or $e^{\mcL} = U(n)$, where $U(n)$ is the Lie group of unitary matrices of dimension $n$. 
The system is also said to be fully controllable when $\text{dim}(\mathcal{L}) = n^2 -1 = \text{dim}(su(n))$, where $su(n)$ is the Lie algebra of matrices of $u(n)$ with zero trace, or equivalently when $\mcL = su(n)$ or $e^{\mcL} = SU(n)$, where $SU(n)$ is the set of matrices of $U(n)$ with determinant equal to one.
This last statement is equivalent to saying that all quantum states are defined up to an unmeasurable global phase. 

\subsection{Ans\"atze for VQAs}\label{sec:ansatze}
The flexibility of VQAs comes notably from the freedom one has when designing the ansatz~$U(\bm\theta)$, used to find approximate solutions.
The success of the algorithm therefore strongly depends on the quality of the ansatz.
In most cases, it is possible to separate gate-based variational forms into two categories: (i) hardware-efficient approaches and (ii) Pauli-based circuits.
The former aims at producing shallow circuits and reducing the use of expensive two-qubit gates by relying on quantum operations (gates) that are native to the processor's architecture~\cite{kandala2017hardware,barkoutsos2018quantum}.
A drawback of this method is that the quantum circuits seem random from the problem Hamiltonian standpoint.
Consequently, their optimization landscape is plagued by the~\textit{barren plateaus} problem~\cite{McClean_2018}: the average gradient magnitude approaches zero for circuits of even modest parameter scaling with respect to the number of qubits.

In this article, we are interested in quantum circuits that are generated by unitaries resulting from the exponentiation of a \textit{circuit} Hamiltonian.
The circuit Hamiltonian~$\hH_c$ can generally be written as a linear combination of the Pauli strings
\begin{equation}
\hH_c = \sum_\alpha c_\alpha \hat P_\alpha, \quad c_\alpha \in \mathbb{C},
\end{equation}
where the~$\hat P_\alpha$ are~$n$-qubit Pauli strings of the form
\begin{equation}
    \hat P_\alpha = \bigotimes_{j} \sigma_a^{(j)},
\end{equation}
with~$\sigma_a^{(j)}\in \{\hI,\hX,\hY,\hZ\}$ a Pauli matrix applied on qubit~$j$.
We refer to this class of parameterized quantum circuits as Pauli-based PQCs or P-PQCs.

The chosen circuit Hamiltonian and the resultant circuit thus depend on the target application of the VQA.
For example, in the case of classical combinatorial optimization with the Quantum Alternating Operator Ansatz (QAOA)~\cite{hadfield2019quantum},~$\hH_c = \hH_p + \hH_m$ where~$\hH_p$ is a diagonal phase Hamiltonian and~$\hH_m$ is a mixing Hamiltonian that preserves symmetries of the solution space, \eg, the number of excitations.
For the Variational Hamiltonian Ansatz (VHA)~\cite{wecker2015progress},~$\hH_c = \Hp$ is directly the problem Hamiltonian whose ground state we wish to prepare.
In the QOCA approach~\cite{choquette2020quantum},~$\hH_c = \Hp+\sum_k \hH_k$, where~$\{\hH_k\}$ is a collection of \textit{drive} Hamiltonians whose purpose is to improve convergence by slightly breaking symmetries of~$\Hp$.  
Finally, in the Unitary Coupled Cluster (UCC) ansatz for quantum chemistry,~$\hH_c$ implements a set of single- and double-excitation operators~\cite{peruzzo2014variational}.
We discuss the construction of a PQC from~$\hH_c$ in the next section.

\section{Method}\label{sec:method}

\subsection{Controllability and Pauli-based PQCs}
A Pauli-based PQC is derived from the circuit Hamiltonian by first splitting~$\hH_c = \sum_i \hH_i$ into groups of Hamiltonian terms~$\hH_i$ and then associating a variational parameter~$\theta_i$ to each of them.
We then take the exponential of~$\hH_c$ and perform a first-order Trotter-Suzuki decomposition~$e^{A+B}\sim e^A e^B$ to obtain a product of unitaries.
For a particular splitting, a~$p$-layer quantum circuit therefore reads
\begin{equation}\label{eq: P-PQC}
        U(\bm\theta) = \prod_{d=1}^p \prod_{j} e^{i\theta_{d,j}\hH_j},
\end{equation}
where~$\bm\theta = \{\theta_{d,j}\}$ collects all variational parameters.
To implement individual~$e^{i\theta_{d,i}\hH_i}$ as a circuit, a similar decomposition as in~\cref{eq: P-PQC} is often used. 
Note that a general procedure to construct circuits corresponding to exponentials of Pauli strings, as it is the case here, was introduced in Ref.~\cite{whitfield2011simulation}.
The finer details of circuit compilation can be found in Refs.~\cite{anand2022quantum} and \cite{kottmann2021tequila}.

The partitioning of~$\hH_c$ determines the position of the variational parameters in~$U(\bm\theta)$.
Therefore, we refer to a given partition as a \textbf{parametrization} of the ansatz. 
Observe that the number of possible parameterizations is combinatorially large and it is far from trivial to establish~\textit{a priori} which one may yield the better results~\cite{izmaylov2020order, grimsley2019trotterized}.

To define controllability of Pauli-based PQCs, we again turn to Theorem 1.
We note that the mathematical form of P-PQCs~\eqref{eq: P-PQC} resembles the exponential map~$e^{\mathcal{L}}$ of the Lie algebra~$\mathcal{L}$ associated with the circuit Hamiltonian~$\hH_c$ defined in~\cref{sec:ansatze}.
In short,
\begin{align*}
    \hH (t)                &\longrightarrow \hH_c, \\
    e^{\mcL}               &\rightsquigarrow \prod_{d=1}^p \prod_{j} e^{i\theta_{d,j}\hH_j}.
\end{align*}
The squiggly arrow connecting~$e^{\mcL}$ to the P-PQC's circuit equation denotes that the proposed layered ansatz approach does not exactly implement the exponential map~$e^{\mcL}$.
The latter is true only in the limit of an infinite number of layers, where the first-order Trotter-Suzuki approximation is exact.
This observation allows us to make a connection with the theory of controllability of quantum systems, which provides tools that will serve useful to gain insight into the performance of different parameterizations.

\subsection{Reachability of PQCs}
We first present a method to generate the set of reachable states for a given Hamiltonian~$\hH(t)$ in \cref{alg: Lie algebra}.
It is an iterative procedure, where a basis for the dynamical Lie algebra of the system is generated using all the commutators of elements of~$\mcL$ and appending~$\mcL$ whenever a new linearly independent operator is found. 
\cref{alg: Lie algebra} is analogous to generating a vector space basis using the Gram-Schmidt orthogonalization method~\cite{d2007introduction}. 

\begin{algorithm}[htbp!]
\SetAlgoLined
\textbf{Input}: controlled terms of~$\hH(t)$\\
\textbf{Output}: Dynamical Lie algebra\\
\textbf{Initialize}: current Lie algebra~$\mcL_{\text{current}}$ with all controlled terms of~$\hH(t)$\;
 1. Generate candidates $C\equiv\{ [A_\alpha, A_\beta]: A_\alpha, A_\beta\in \mcL_{\text{current}}\}$\;
 2. Find the set of elements $I$ in~$C$ that are linearly independent from the elements of~$\mcL_{\text{current}}$\;
 \While{$I$ is non-empty and dim($\mcL_{current}$) < ~$n^2 -1$}{
    a. Add $I$ to~$\mcL_{\text{current}}$\;
    b. Generate candidates~$C\equiv\{ [A_\alpha, A_\beta]: A_\alpha, A_\beta\in \mcL_{\text{current}}\}$\;
    c. Find the set of elements $I$ in~$C$ that are linearly independent from the elements of~$\mcL_{\text{current}}$\;
 }
 \textbf{return}: $\mcL_{current}$
 \caption{\label{alg: Lie algebra}Generating a basis of the dynamical Lie algebra}
\end{algorithm}

Following this procedure in \cref{alg: Lie algebra},~$\text{dim}(\mathcal{L})$, or the~\textbf{Lie rank criterion}, is obtained by simply counting the number of elements in the final basis generated by the algorithm.
In brief, the set of reachable states of a controlled system~$\hH(t)=\sum_k c_k(t)\hH_k$ depends on which Hamiltonian terms~$\hH_k$ are associated with control functions~$c_k(t)$.
For example, the Lie algebra of~$\hH_1 = a(t)\hX + b(t)\hZ$ will be of higher dimension than that of~$\hH_2 = c(t)(\hX + \hZ)$. 
In fact,~$\hH_1$ is fully controllable, while~$\hH_2$ is not.

The procedure to determine the set of reachable states of a given P-PQC is, therefore, the same as in \cref{alg: Lie algebra}, where the control Hamiltonian~$\hH(t)$ is replaced by~$\hH_c$.
Further, the initial Lie algebra is now given by the set of all terms that are parametrized in the variational ansatz, that is the set~$\{\hH_i\}$ from~\cref{eq: P-PQC}.
By running this procedure, we can obtain the dimension of the solution space, which is a measure of the subspace of the full Hilbert space one can wish to reach with a given parametrization of the P-PQC.
A one-qubit example of the computation of the dynamical Lie algebra is presented in~\cref{app: one qubit}.
In our opinion, this is a very powerful tool for the design of ans\"atze as one can predict which coverage of the Hilbert space will be explored through the variational optimization.

\section{Results and Discussion}\label{sec:simulations}

We now build upon these concepts by studying practical applications of VQAs through the lens of dynamical Lie algebras.
First, using a generic two-qubit Pauli-based ansatz, in~\cref{sec: scaling with params} we show evidence of the scaling of the Lie rank criterion as a function of the number of parameters in the ansatz.
Furthermore, we analyze the performance of different parametrizations of Pauli-based ans\"atze applied to the XXZ-Heisenberg model in~\cref{sec: heisenberg}.
In addition, we correlate our results with the Lie algebra's dimension of the ans\"atze found using \cref{alg: Lie algebra}.

\subsection{Relationship between the dimension of the solution space with the number of parameters}
\label{sec: scaling with params}

\begin{figure}[htbp]
        \begin{center}
        \includegraphics[width=0.99\columnwidth]{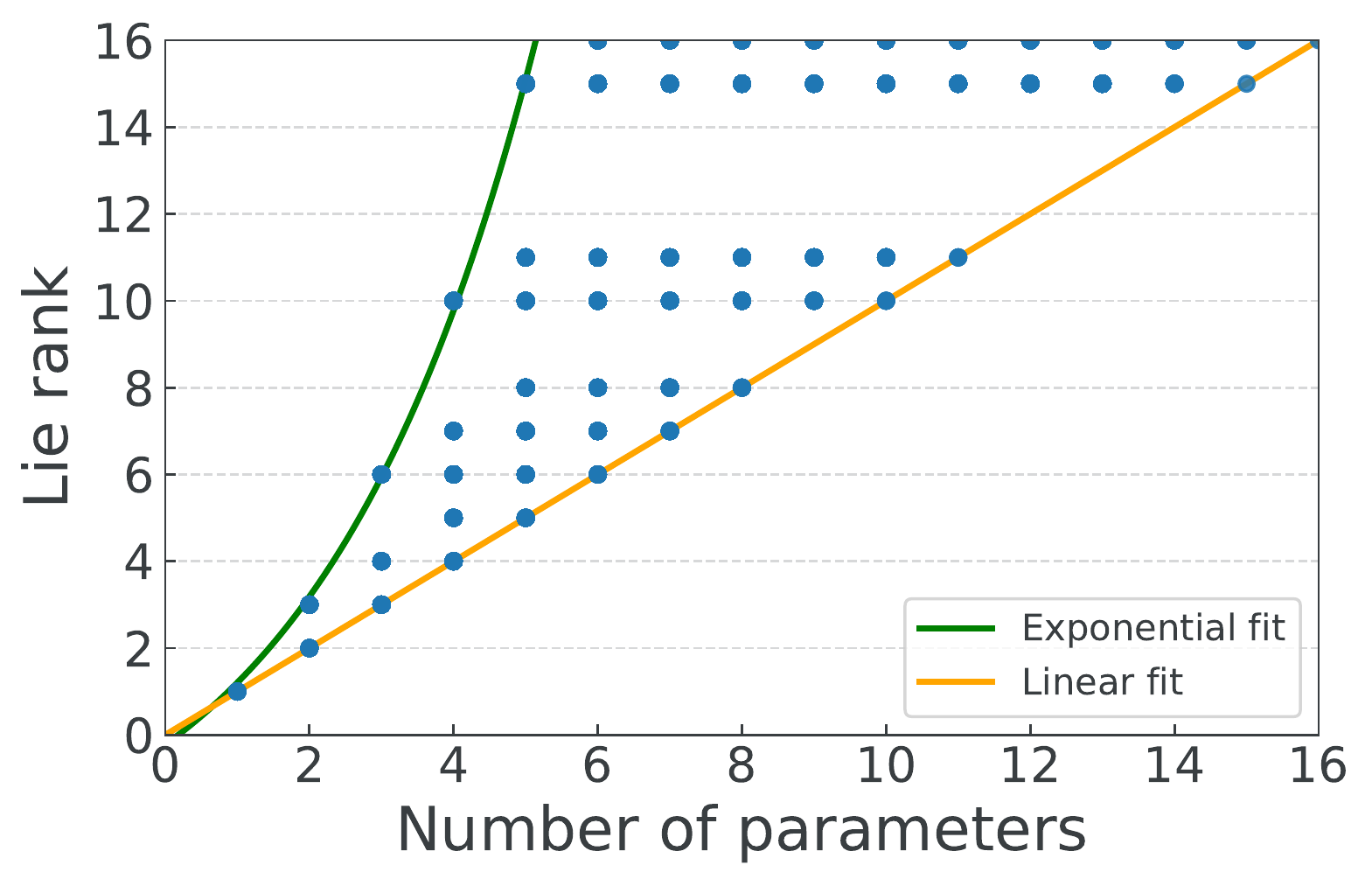}
        \caption{\label{fig: scaling}
        Value of the Lie rank criterion defined by the dimension of the dynamical Lie algebra of two-qubit circuit Hamiltonians of varying number of parametrized terms.
        For~$m$ number of parameters, the~${16 \choose m}$ possible combinations of two-qubit Pauli strings were considered.}
        \end{center}
\end{figure}

We study the scaling of the Lie rank criterion as a function of the number of parametrized Pauli terms in the P-PQC.
For this task, we consider the~$n=2$ qubit Pauli algebra defined by the set~$\{\hat\sigma_a\otimes\hat\sigma_b\}$, where $\hat\sigma_a,\hat\sigma_b \in \{\hI,\hX,\hY,\hZ\}$, to design the ansatz.
To build a circuit Hamiltonian with~$m$ variational parameters, we consider the~${16 \choose m}$ ways to sample~$m$ Pauli strings from the~$4^2=16$ total elements of the Pauli group.
With~$m$ ranging from 1 to 16, this represents precisely 65,535 instances.
For all~$m$, we compute the dynamical Lie algebra of the~${16 \choose m}$ parametrized terms and report the final Lie rank criterion value in~\cref{fig: scaling}.

We observe that full Hilbert space coverage is not possible with fewer than five parametrized terms and is guaranteed with 15 and up.
Indeed, as discussed previously, a system is fully controllable when~$\dim(\mcL)= 4^n-1$, hence 15 in our case.
Therefore, these results entail that there exists ans\"atze with $5\geq m \leq 16$ parameters that are fully controllable in the limit of large number of layers.

Next, we note that the scaling of the Lie rank criterion with the number of parameters is at worst linear and at best exponential, depending on the parametrization.
This means that a clever choice of Pauli terms to include in the circuit Hamiltonian can offer a favorable coverage of the Hilbert space for a minimal number of variational parameters.
One may also utilize methods such as ADAPT-VQE \cite{grimsley2019adaptive, tang2021qubit} to adaptively find the optimal parametrization.

We now present the XXZ-Heisenberg model and report our findings from the analysis of the different parameterizations using \cref{alg: Lie algebra}.

\subsection{Heisenberg model}
\label{sec: heisenberg}
The Heisenberg model is a widely used statistical mechanical model that describes a chain of atoms with nearest neighbour interactions. The Hamiltonian of such a system on a two-dimensional lattice can be written as follows:

\begin{equation}
    \hat{H} = \sum_{\langle i,j\rangle} \left(  - J_X \hat{X}_{i} \hat{X}_{j} - J_Y \hat{Y}_{i} \hat{Y}_{j} - J_Z \hat{Z}_{i} \hat{Z}_{j} \right) - h \sum_{k} \hat{Z}_k 
\end{equation}
where $\hX, \hY , \hZ$ are the Pauli matrices, $\langle i,j\rangle$ denotes all the pairs of adjacent lattice sites, and $J_X, J_Y, J_Z$ are the coupling constants and $h$ represents the external magnetic field. 
The values of the coupling constant determine the model that is used to describe the system, for $ J = J_X = J_Y \ne J_Z \equiv \Delta$, the model is known as the XXZ-Heisenberg model. 

In this section, we report results from VQA simulations for estimating the ferromagnetic ground state energy of the $2\times 2$ XXZ-Heisenberg model, with $\Delta/J = -20$.
The resulting Hamiltonian is composed of 13 different 4-qubit Pauli terms.
The ansatz we choose for the VQA is the Variational Hamiltonian Ansatz (VHA)~\cite{wecker2015progress}, which builds P-PQCs sampling Pauli terms from the problem Hamiltonian.

\subsubsection{Lie rank scaling vs parameterization}\label{sec:lrvspar}
\begin{figure}[htbp]
\centering
{\includegraphics[width=0.9\columnwidth]{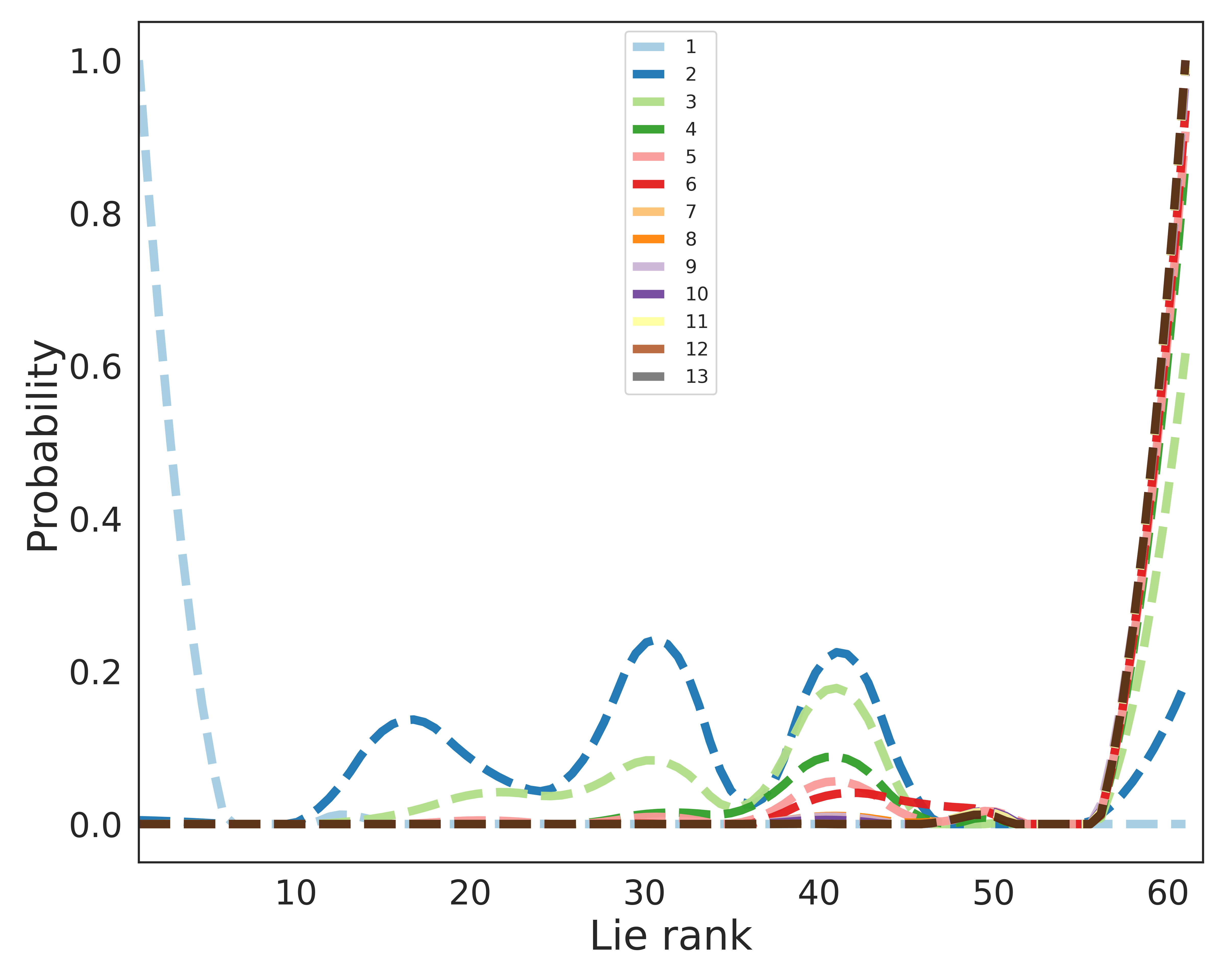}}\\
{\includegraphics[width=0.9\columnwidth]{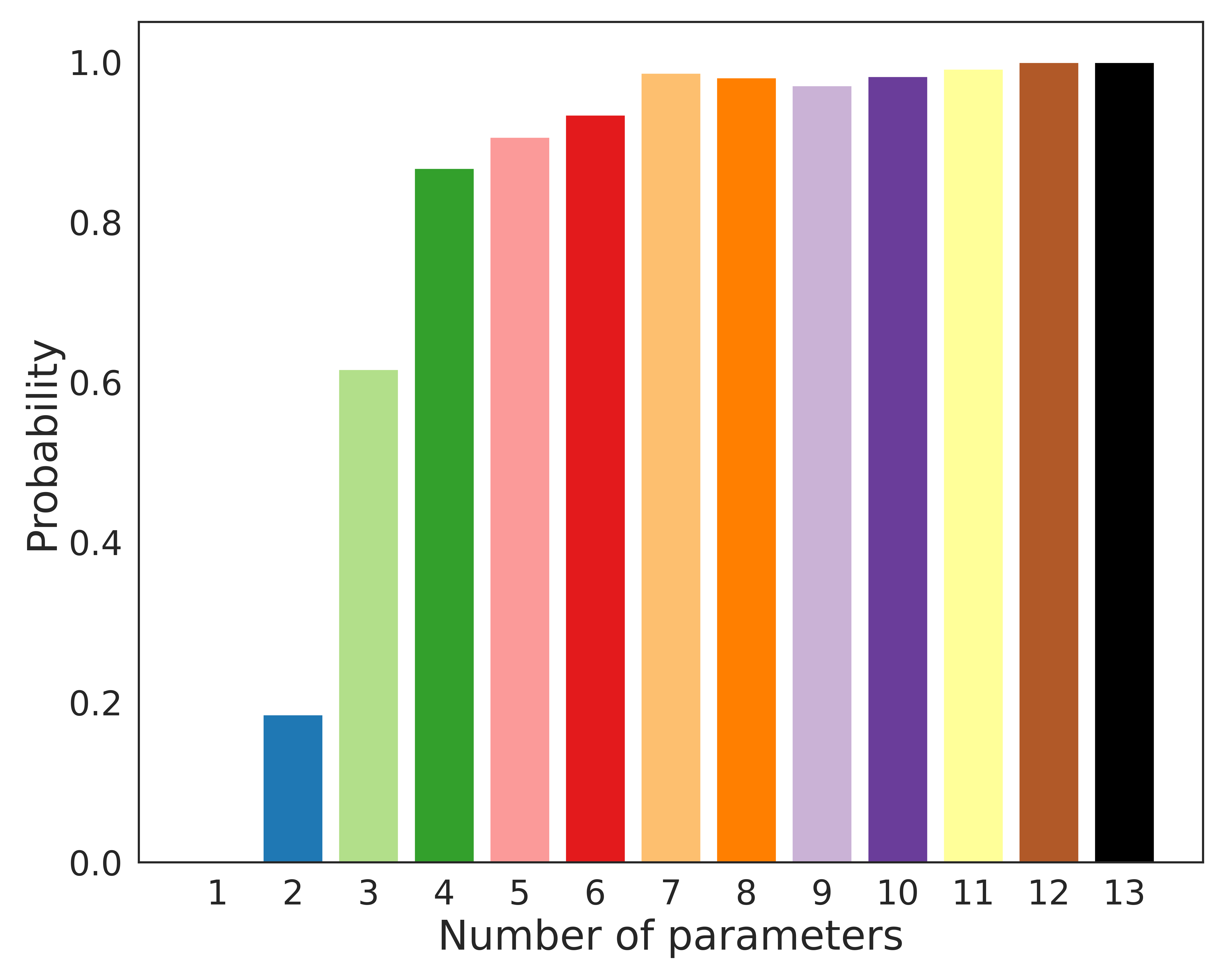}}\\
\caption{(Top) Probability distribution of the calculated Lie rank of different $m$-parameter partitions of the XXZ-Heisenberg Hamiltonian for $m$ ranging from 1 to 13. (Bottom) Probability of achieving the maximum possible Lie rank (61) of the VHA ansatz using different parameterizations of the Hamiltonian.}
\label{fig:lrvspar}
\end{figure}

We calculate the Lie rank criterion of different partitions of the XXZ-Heisenberg Hamiltonian terms.
We do this by choosing a number $m$ ranging from 1 to 13 and dividing the Hamiltonian terms into $m$ independent sets, each set receiving a controllable parameter.
Each such way of dividing up the Hamiltonian into $m$ sets constitute a  partition.
For all $m$, we then uniformly sample a partition from all the possible $m$-parameter partitions and calculate the Lie rank using \cref{alg: Lie algebra}.
We use the results from the different runs to generate a probability of having a particular Lie rank given the number $m$.
The distribution is generated by considering all the Lie rank values for the different partitions and then normalizing it as:
\begin{equation}
    P(l_r|m) = \frac{n_{l_r}}{n_T},
\end{equation}
where, $P(l_r|m)$ represents the probability of having a Lie rank $l_r$ given $m$-set partitions of the Hamiltonian, $n_{l_r}$ is the number of time a $m$-set partition of the Hamiltonian had a Lie rank $l_r$, and $n_T$ is the total number of $m$-set partitions of the Hamiltonian considered.
The distribution of Lie rank for the different number of partitions $m$ is plotted in ~\cref{fig:lrvspar}.

We first note that full Hilbert space coverage, which would require a Lie rank of $4^4-1=255$, is not possible in this case since the maximum Lie rank calculated is 61.
This means that the VHA, even when fully parametrized, is far from being expressible enough to cover the full Hilbert space.


From \cref{fig:lrvspar} (top), we observe that for a single controllable parameter (partition) the probability distribution of Lie ranks is centered around 1.
However, as we increase the number of partitions in the circuit Hamiltonian, this distribution shifts away from 1 to eventually re-center around 61, the maximum Lie rank one can achieve with the VHA.
This is the expected behavior, as increasing the number of partitions will lead to increased controllability of the system.

\cref{fig:lrvspar} (bottom) indicates that the maximum Lie rank is systematically obtained when using $\geq 12$ parameters or in some instances comprising of as low as 2 parameters.
Also, with only 4 controllable parameters the probability of reaching the largest Lie rank is already greater than 0.8.
This means that in principle, 4 parameters are enough to achieve a similar Hilbert space coverage than that of a 13-parameter ansatz.
However, in practice, one might have to consider more layers of the ansatz in the 4-parameter case as these estimates hold exactly only in the large number of layers limit.
Moreover, knowing with certainty that a given 4-parameter partition leads to a high Lie rank might require more iterations of \cref{alg: Lie algebra} than a 13-parameter partition, which requires 1 iteration.
We investigate this next by looking at the evolution of the Lie rank within \cref{alg: Lie algebra} for all the partitions considered here.

\begin{figure*}[htbp]
        \begin{center}
        \includegraphics[width=0.9\linewidth]{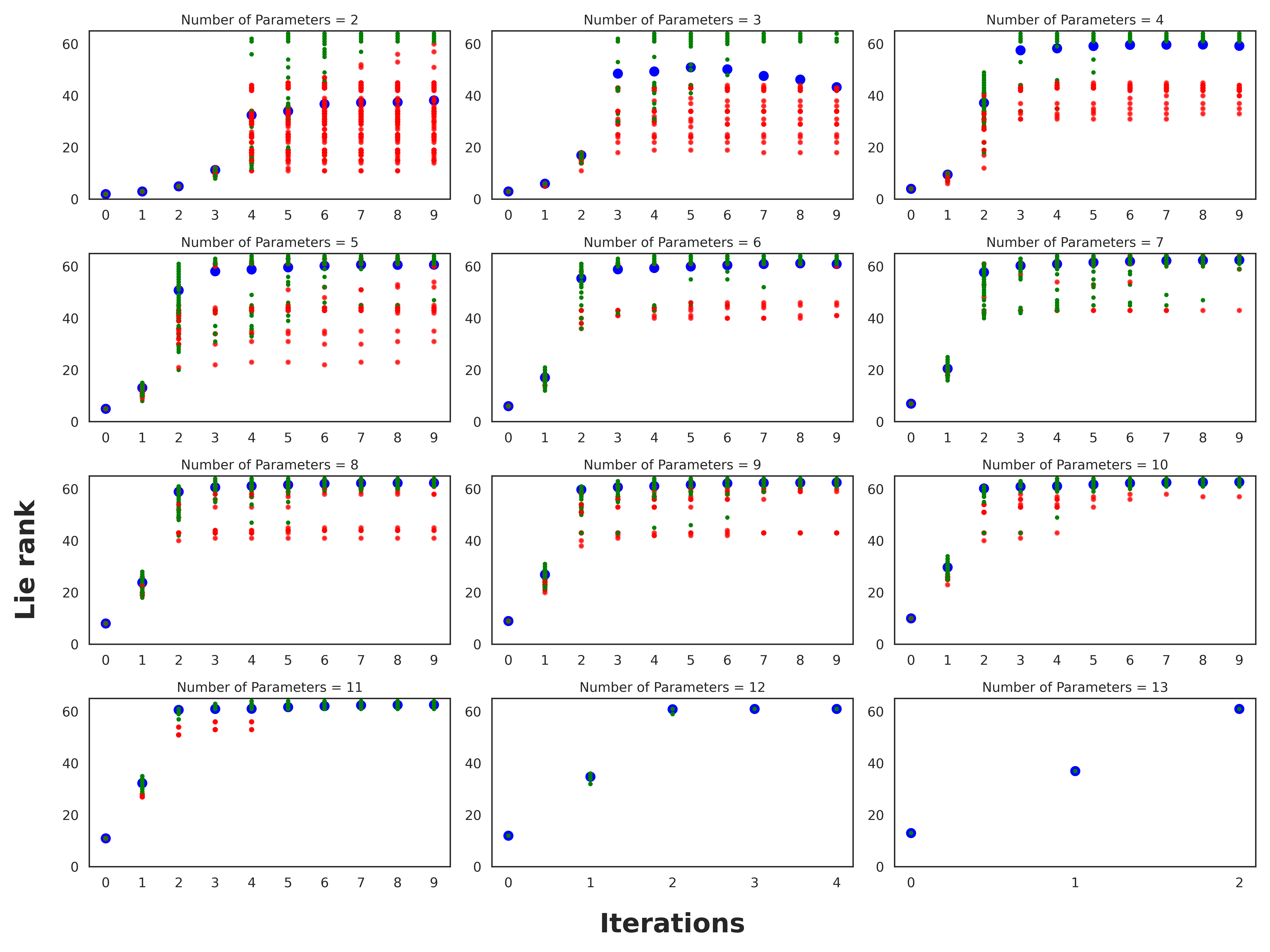}
        \caption{\label{fig:lrev_full} Calculated Lie rank as a function of the order of commutators used (iterations of \ref{alg: Lie algebra}) for different parameterization of the VHA ansatz. 
        The blue dots represent the mean Lie rank at a particular iteration, and the data in red and green represent the particular parameterizations that eventually reached (green) or not (red) the maximum Lie rank of 61. 
        Note that for all plots the number of iterations has been truncated at 9 for clarity.}
        \end{center}
\end{figure*}

\subsubsection{Evolution of the Lie rank in \cref{alg: Lie algebra}}\label{sec:lrev}
In \cref{alg: Lie algebra}, the Lie rank criterion is calculated by counting the number of linearly independent candidates that can be generated using higher-order commutators of the initial partition of the Hamiltonian.
We analyze the evolution of the Lie rank as a function of the number of iterations of \cref{alg: Lie algebra}, \ie the order of the commutators for the different parameterizations of the XXZ-Hamiltonian considered in~\cref{sec:lrvspar}, and plot them in~\cref{fig:lrev_full}.
It can be seen from the plots that the mean Lie rank for the different parameterization increases logistically.
However, with an increasing number of controllable parameters in the circuit Hamiltonian, the point of inflection shifts to a lower value in iterations.
This implies that the higher the number of partitions, the higher the probability of obtaining a large coverage with a low number of repetitions is.

We now make some observations about the growth of the Lie rank value for the different parameterizations considered above.
\begin{enumerate}
    \item The parameterizations which can reach the maximum achievable rank (represented by the green data in the plot) achieves it with a lower order of the commutators (iterations).
    That is the mean Lie rank value is very close to the maximum achievable value within the first 5 iterations for a majority of the parameterizations we considered.
    \item The parameterizations (3 or more parameters) which have a Lie rank value greater than the mean Lie rank at any iteration mostly achieve the maximum Lie rank value.
    \item The probability of reaching the maximum achievable Lie rank is greater than 80 percent (see \cref{fig:lrvspar} (bottom)) for a majority of parameterizations considered here.
\end{enumerate}

\begin{figure}[htbp]
        \begin{center}
        \includegraphics[width=0.85\columnwidth]{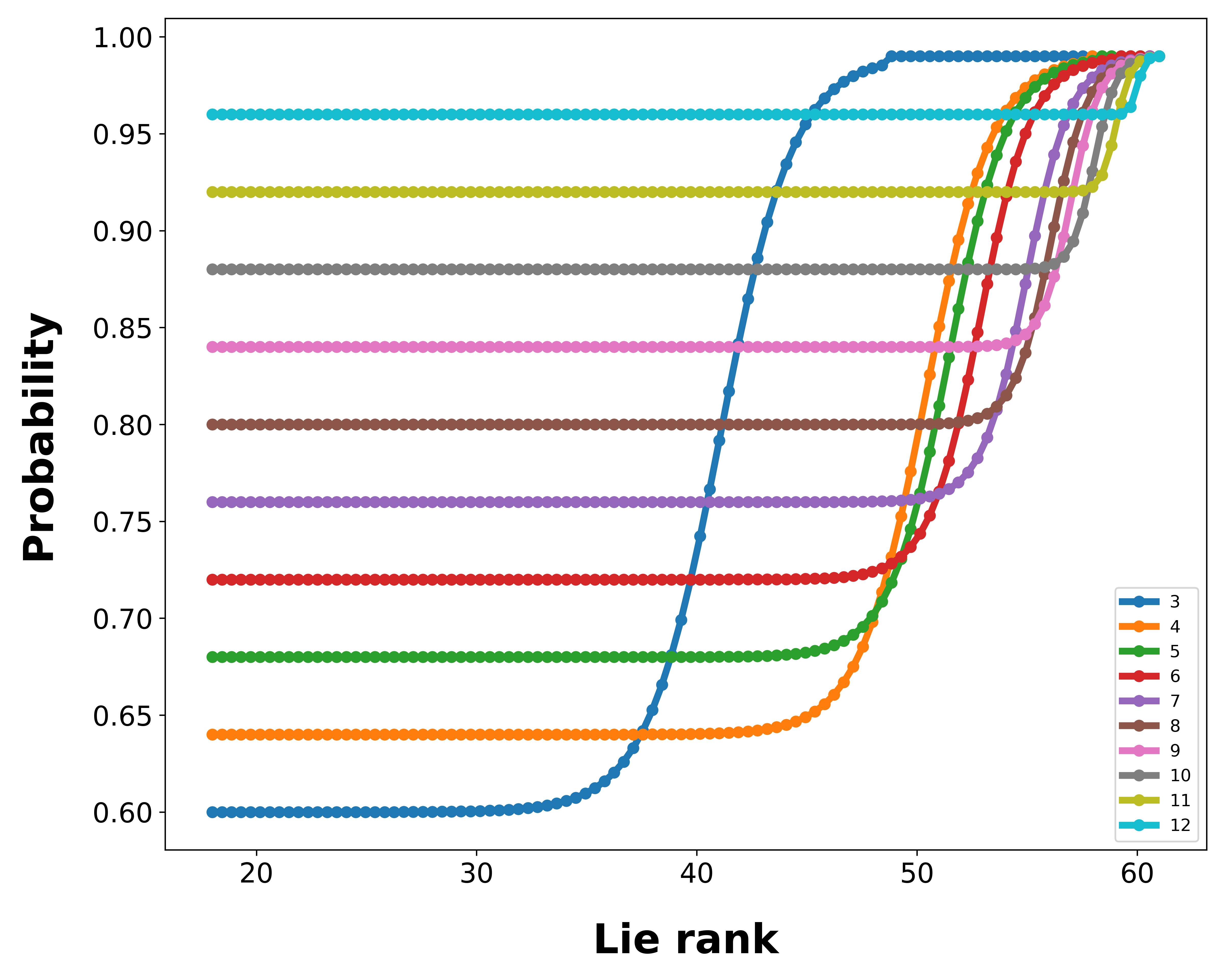}
        \caption{\label{fig:lrev_}Probability of \cref{alg: Lie algebra} outputting the maximum Lie rank as a function of the Lie rank after three iterations of the algorithm. 
        The different colored lines represent the number of partitions in the Hamiltonian for the VHA ansatz.
        }
        \end{center}
\end{figure}

Based on these observations, we can define a function $L(x)$ that approximates the probability of reaching the maximum achievable Lie rank, given a particular Lie rank value at some iteration. 
We first define a function $f(x)$,
\begin{align}
    f(x) =\begin{cases}
        1 \quad x \geq a, \\
    \frac{1}{1 + \alpha e^{\beta x}} \quad \text{else},
    \end{cases}
\end{align}
where, $a$ is the mean Lie rank value at the given iteration, $\alpha$ and $\beta$ are some constants that control the shape of the curve.
The approximate probability function $L(x)$ is then defined as a one-dimensional linear interpolation of the function $f(x)$.
The probability distribution $L(x)$ with third-order commutators for the different number of partitions is plotted in \cref{fig:lrev_}.
It can be seen that using this distribution we can approximately predict the reachability of a given parameterization without running the full \cref{alg: Lie algebra}.

We note that to construct the probability distribution, one needs an estimate of the mean Lie rank value at a given iteration and the minimum probability of reaching the maximum rank.
We can estimate the mean Lie rank by running different experiments at a given iteration.
The minimum probability, however, is not available without running the full \cref{alg: Lie algebra}. 
But based on our numerical experiments, we suggest using a small value (0.25) for the smaller number of partitions and linearly increasing it to a higher value (0.95) as we reach the maximum number of parameters.
This approximate distribution can be important for larger problems, as we note that the computation necessary to run the full \cref{alg: Lie algebra} is not scalable.

\subsubsection{Energy vs parameterization}
We have analyzed the coverage of Hilbert space spanned for a given parameterization of a circuit Hamiltonian until now.
Here, we report the results from the simulations for estimating the ground state energy of the XXZ-Heisenberg model for different parameterizations of the Hamiltonian.
We first use an ansatz with a circuit Hamiltonian containing all the terms in the Lie algebra of fully parameterized XXZ-Hamiltonian and calculate the ground state energy.
The ansatz, referred to as the Lie Algebra Partition (LAP) ansatz, contains 61 parameters (or terms), and is constructed as per \cref{eq: P-PQC}.
In theory, this ansatz represents the most expressible PQC one can construct using the Lie algebra of the problem Hamiltonian.
In other words, one layer of the LAP should be as expressible as a sufficiently-parametrized VHA in the limit of large number of ansatz layers.
The final optimized energy with the LAP ansatz is -1.0721, which is very far away from the actual ground state energy (-1.9794) of the Hamiltonian. 
This is indicative of the low reachability of the ansatz, as even with 61 parameters, the ansatz does not have access to the state space corresponding to the ground state of the system.

\begin{figure}[htbp!]
        \begin{center}
        \includegraphics[width=0.99\columnwidth]{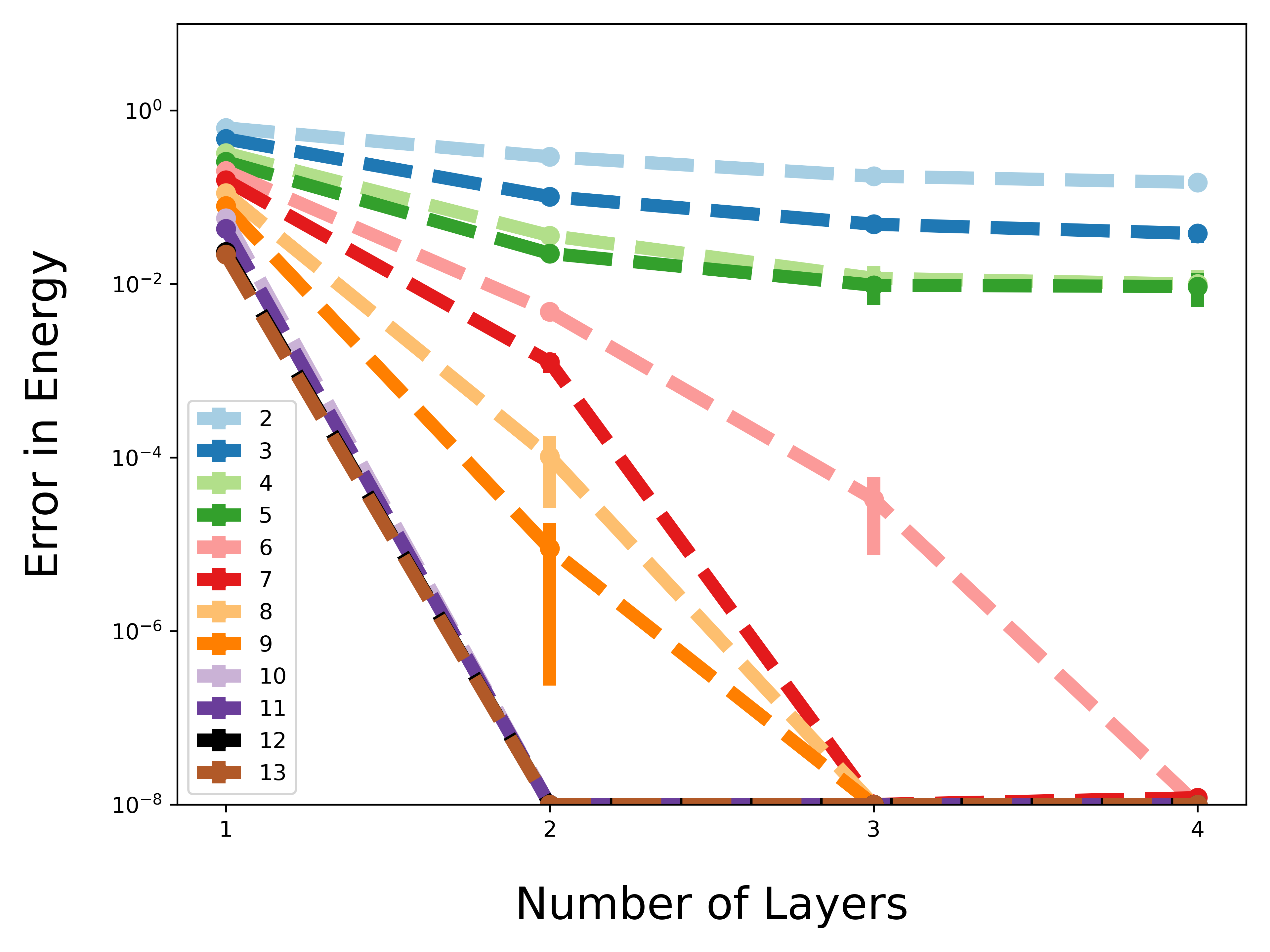}
        \caption{\label{fig:enervspar}Error between the final energy of the VHA ansatz (with different number of parameters) and that of the LAP ansatz (61 parameters) as a function of the number of layers of the VHA ansatz used in the VQE optimization. 
        For each number of parameters, multiple VQE runs with different ways of parametrizing the ansatz are averaged.
        The solid dots represent the mean error value and the vertical lines represent one standard deviation in the error corresponding to different VQE runs. 
        }
        \end{center}
\end{figure}

We next calculate the ground state energy with VHA using an increasing number of layers of the circuit corresponding to different parameterizations used in  \cref{sec:lrvspar} and \cref{sec:lrev}.
The results are presented in \cref{fig:enervspar}.
We plot the error between the final energy of the VHA and the final energy of one layer of the optimized LAP ansatz.
For every number of variational parameters, adding layers brings the results closer in energy to that of the LAP ansatz.
This is expected since the addition of ansatz layers increases the number of parameters, therefore making the circuits more expressible.
However, from the data we notice that no instance of the VHA provides an energy lower than the one obtained optimizing the LAP ansatz.
This confirms that the lowest possible energy achievable using a VHA ansatz corresponds to the LAP ansatz.
However, using directly the LAP is not scalable as its circuit depth is exponentially long.

Luckily, there seems to be a threshold in the number of parameters for which the VHA attains its full expressibility much faster.
Indeed, all instances of the VHA with $\leq 5$ variational parameters show a very slow convergence towards the LAP energy while instances with $\geq 6$ parameters reach the LAP energy within $10^{-8}$ of error with a small number of circuit layers.
This means that one needs not to use the unscalable LAP ansatz in practice.
However, we note that when computationally feasible, optimizing the LAP ansatz is a powerful tool as it gives a lower bound on the performance of the VHA, or any P-PQC-based ans\"atze more generally.

Finally, we note that if the partition of the Hamiltonian is not expressive enough increasing the number of layers will not help in reducing the error in the energy estimation. 
One should rather focus on increasing the reachability of the ansatz.
One such approach is introducing extra terms in the circuit Hamiltonian that break symmetries of the problem Hamiltonian~\cite{choquette2020quantum,vogt2020preparing}, therefore expanding the dynamical Lie algebra of the circuit Hamiltonian.
In these methods, the goal is to cover a larger portion of the Hilbert by introducing drive terms that do not commute with the problem Hamiltonian.
We point out that using such an approach can lead to better ground state energies as the Lie rank increases drastically.
We deduce thereby that one has to keep in mind the dynamic Lie rank of the partition while designing an ansatz - the higher the rank, the higher the chances of the ansatz reaching the correct ground state. 

\section{Conclusion}\label{sec:conclusion}
In this article, we have proposed a new tool for analyzing the expressibility of Pauli-based quantum circuits.
We do this by employing a connection between VQAs and quantum controllability and present an iterative procedure to quantify the reachability of a parametrized ansatz.
We have detailed preliminary explorations on how the choice of parametrization in variational Hamiltonian ans\"{a}tze affects the expressive capability of such ans\"{a}tze using the XXZ-Heisenberg model.

Our procedure is as follows: We first calculate the Lie rank criterion for different parameterizations of a Hamiltonian.
Through numerical experiments, we find that the Lie rank criterion of the Hamiltonian that defines the system ansatz positively correlates with the ability of the Pauli-based ansatz to approximate ground-state energies. 
We also put forward a way to approximate the final Lie rank of a given parametrization using only a few iterations of our costly \cref{alg: Lie algebra}.
The approximation however uses some heuristics and requires the knowledge of the maximum Lie rank.

Finally, we note that the algorithm we propose to generate the Lie rank criterion has an exponential scaling in the number of qubits, making its use computationally intensive, even for the relatively modest systems analyzed in this work.
Thus, for the Lie rank criterion to be a salient design criterion, efficient but accurate algorithms that approximate the Lie rank criterion will be necessary. 

This work adds to the growing efforts towards using tools from quantum control to determine ansatz suitability for different applications, such as in quantum chemistry and classical combinatorial optimization. 
A lot needs to be done before this can be used readily, such as establishing systematic connections between Lie rank and methods for partitioning the problem Hamiltonian to make an informed decision when choosing an ansatz.
Some other areas include extending the framework to non-problem-inspired P-PQCs and using the Lie algebra criteria to choose drive terms similar to gradient-based adaptive ans\"atze construction~\cite{grimsley2019adaptive, ryabinkin2018qubit, ryabinkin2020iterative, tang2021qubit}.
We leave this for future research and hope the community finds ways to extend the presented framework for the aforementioned problems.

\section*{Acknowledgements}
The authors thank Jakob S. Kottmann, Philipp Schleich, Lasse B. Kristensen and Mohsen Bagherimehrab for providing valuable comments regarding the manuscript.
A.A.-G. acknowledges the generous support from Google, Inc.  in the form of a Google Focused Award.
A.A.-G. also acknowledges support from the Canada Industrial Research Chairs  Program and the Canada 150 Research Chairs Program. Computations were performed on the niagara supercomputer at the SciNet HPC Consortium~\cite{niagara1, niagara2}. SciNet is funded by: the Canada Foundation for Innovation; the Government of Ontario; Ontario Research Fund - Research Excellence; and the University of Toronto.

\bibliography{biblio.bib}

\appendix
\section{Dynamical Lie algebra: a one-qubit example}
\label{app: one qubit}
For this example, let the circuit Hamiltonian be the $\hX$ Pauli matrix
\begin{equation}
    \hH_c=\hX.
\end{equation}
The dynamical Lie algebra is~$\mcL = \{i\hX\}$, which has a dimension of~$\dim \mcL = 1 < n^2-1=3$, where $n=2$ is the size of the Hilbert space.
This means that a P-PQC of the form
\begin{equation}\label{eq: eiX}
        U(\theta) = e^{i\theta\hX}
\end{equation}
can not reach all possible one-qubit states.
This is obvious since~\cref{eq: eiX} implements a rotation above the~$x$ axis of the Bloch sphere and the set of reachable states form a ring around this axis.

Now, let~$\hH_c=\hX+\hY$ and~$U(\theta_x,\theta_y) = e^{i\theta_x\hX}e^{i\theta_y\hY}$.
To find the dynamical Lie algebra, we must first take the commutator between terms of~$\hH_c$ to find
\begin{equation}
        [\hX,\hY] \sim \hZ.
\end{equation}
$\hZ$ is a new matrix that is linearly independent of $\hX$ and $\hY$ and we therefore append it to the dynamical Lie algebra of the system.
Note that taking the commutator of $\hZ$ with terms of $\hH_c$ does not allow to find additional linearly independent matrices.
The dynamical Lie algebra is thus~$\mcL = \{i\hX,i\hY,i\hZ\}$, which has a dimension of~$\dim \mcL = 3 = n^2-1$.
The system~$\hH_c=\hX+\hY$ is therefore said to be fully controllable as all one-qubit states can be obtained by tuning~$\{\theta_x,\theta_y\}$ appropriately.


\end{document}